\documentclass[sts]{imsart}

\usepackage{algorithm,algorithmic,amsfonts,amssymb,amsmath,amscd,amsthm,bm,latexsym}
\usepackage{natbib}
\usepackage{graphicx}
\usepackage{picins}
\usepackage{upgreek}
\usepackage{units}
\usepackage{nicefrac}
\usepackage{comment}

\renewcommand{\rho}{\varrho}

\begin{document}

\begin{frontmatter}

\title{Nonparametric Bayesian clay for robust decision bricks}
\runtitle{Discussion of Approximate models and robust decisions}
\thankstext{T1}{Christian P. Robert and Judith Rousseau, CEREMADE, Universit{\' e} Paris-Dauphine, 
PSL, 75775 Paris cedex 16, France
{\sf xian,rousseau@ceremade.dauphine.fr}. Research partly supported by the Agence Nationale de la Recherche (ANR,
212, rue de Bercy 75012 Paris) through the 2012--2015 grant ANR-11-BS01-0010 ``Calibration''.
Both authors are members of Laboratoire de Statistique, CREST, Paris. C.P.~Robert is also affiliated as a part-time
professor in the Department of Statistics of the University of Warwick, Coventry, UK.}

\begin{aug}
\author{\snm{Christian P.~Robert and Judith Rousseau}}
\affiliation{Universit{\'e} Paris-Dauphine, PSL, CEREMADE, and CREST, Paris}
\end{aug}

\begin{abstract}
This note discusses  Watson and Holmes (2016) and their proposals towards more robust Bayesian decisions.
While we acknowledge and commend the authors for setting new and all-encompassing principles of Bayesian robustness, and
we appreciate the strong anchoring of those within a decision-theoretic referential, we remain uncertain as to which extent
such principles can be applied outside binary decisions. We also wonder at the ultimate relevance of Kullback-Leibler
neighbourhoods to characterise robustness and favour extensions along non-parametric axes.
\end{abstract}

\end{frontmatter}

\section{Introduction: There is nothing like first-hand evidence}

The most sensitive aspect of both Bayesian and non-Bayesian statistics certainly is the reliance on a 
probabilistic model since even Bayesian non-parametrics relies on a highly concentrated modelling in a
functional space. Considering the issue of misspecification from within Bayesian analysis is therefore a major undertaking that has been curiously overlooked in the past, to the point that it shows up in negative like a missing link in the field. Assessing
the goodness of fit of a given model or exploring the consequences of working with a misspecified model have been little
studied so far and there certainly is no theoretical or methodological corpus that can be acknowledged as a reference.
Bayesian robustness definitely was a keyword in the 80's \citep{berger:1996} but the field somewhat dwindled along the
year, presumably overtaken [as already pointed out by the
authors] by the MCMC tsunami that allowed for much more ambitious modelling and the incorporation of higher variability
through hierarchical structures and priors \citep{cappe:robert:2000,robert:casella:2010}, although this may constitute a
wishful reinterpretation of the past. Another possible reason is that the complex mathematics involved in the formal
representation of the robustness desiderata quickly got intractable. In any case, Bayesian robustness was mostly concerned with reducing the
impact of the prior modelling, i.e., it was understood in terms of robustness against the prior rather than against the
model. (A different approach is adopted in \citealp{samaniego:2010}, who considers a second type of priors intended to
assess the performances of the original priors, with limited applicability.)

In this paper, the authors aim at providing robust inference against possible misspecifications of  the sampling model also, 
producing empirical and methodological perspectives on the issue. This attempt is most commendable and we hope it will
induce others to enlarge and deepen the work in this direction. In particular, we think there is some degree of urgency
in solving the conundrum of the major ``Big Data" challenge, namely the impossible derivation of a complete statistical
model with high dimension or complex data structures (which differs from the ``tall data" case where the sheer size of
the data hinders standard algorithms, see, e.g., \citealp{bardenet:etal:2014}). We like very much the idea that
robustness should be targeted to a specific decision and thus firmly set within a decision-theoretic framework. The paper
starts with a refreshingly modern review on robustness in this context and then advances two types of propositions
towards evaluating robustness of Bayesian procedures and possibly proposing a more robust procedure. However, we find
the empirical assessment that results from the approach somewhat too tentative and too inconclusive to provide a
guidance to practitioners. It indeed appears that expanding the framework beside binary decisions faces considerable
difficulties. We thus suggest below that more (involved) non-parametric procedures should be developed towards this goal. In
our opinion, one of the most challenging foundations of the paper is the
involvement of the action $a$ in the construction of the reference prior, which does not appear natural or acceptable to
us, because the coherence of the Bayesian perspective does not seem to transfer to this approach.

\section{Game of thorns: Duality between posterior and loss function}

First, we want to point out that, in the specific and important setting of discrete decision spaces, and 
in particular for binary decisions, the propositions of the authors make a lot of sense. We are thus genuinely
curious as to how these ideas could extent to infinite and continuous decision spaces. 
While we are similarly strongly inclined towards a decision-theoretic approach to statistics and hence definitely sympathetic to the
perspective adopted in the paper, we also think one should keep in mind the unfortunately ignored remark by Herman Rubin
\citep{rubin:1987b} that model densities and loss functions are only utilised through the product {\sf prior
x likelihood x loss} in Bayesian decision theory. Hence, in this sense, prior and loss are indistinguishable beyond this
product. This reference is meant to
support the point that both sampling and prior densities {\em and} losses should be assessed via a robustness filter,
rather than solely examining the {\sf prior x likelihoo x loss} term under the robustness magnifying glass. There exist
rationality arguments and the like about the choice of
a loss function \citep{raiffa:1968,degroot:1970,berger:1985}, but since most parameters are a by-product [of possibly strong
relevance] of defining a model, rather than enjoying an existence of their own, it is difficult not to think of the
loss function as being linked to the model (mis-)specification.  In particular, this is why we do not see how ``changing
the likelihood changes the interpretation of the prior" (p.6) is such a ``thorny issue" (p.6). The notion of loss 
robustness is only alluded to in the conclusion of the paper and we hope it can be considered much further in parallel
with the current assessment of the model. Obviously there is no free lunch and it is necessary to accept some of the hypotheses to be able to give meaningful conclusions. Our point is that the approaches seem  to make a lot of sense for decisions belonging to a finite space, but may be less so or at least not so obvious sense in continuous cases.

\section{The Baker Street~irregulars : Are Kullback-Leibler neighbourhoods a pertinent choice to assess robustness? }

The authors build and study semi-local measures of robustness, in the sense that the impact of the decision is evaluated on a neighbourhood of the posterior distribution. As nicely reviewed in their Section 2, this idea is not new and Kullback-Leibler neighbourhoods had already been considered in a series of papers. The key justification in the present paper, for Kullback-Leibler neighbourhoods is in Theorem 4.2, where some notion of coherency is invoked.  

The authors build their approach from three principles. It is hard to disagree with these three principles laid out for
D-open methods. However their approach based on the Kullback-Leibler neighbourhoods raises a few issues. 
The result produced in Theorem 4.1 indicates that the least favourable prior involves the exponential of the
loss: this is not surprising given earlier works by the authors like \cite{bissiri:holmes:walker:2013}. 
What sounds rather confusing while being central to the theme of the
paper is the fact that each possible action $a$ induces a different prior or rather distribution in the parameter space. 
To understand this peculiar incorporation of the decision $a$ in the least favourable model (or in posterior
distribution) means that the least favourable decision is twice {\em a posteriori}: it is indeed
an update after both having observed the data and taken an action $a$. This proposal makes complete sense when
evaluating the action, but it gets difficult to understand its meaning when proposing a new action. In other words, it
seems to run against the grain of Bayesian principles. 
The persistence of the indexing in $a$ of the notions and notations throughout the
paper thus remains quite puzzling to us. (Similar puzzlement is attached with the impact of the Monte Carlo variability on the
final decision, as discussed in Section 4.3.)

Besides, the least favourable posteriors only enjoy an implicit expression since the Lagrange multiplier $\lambda_a( C ) $
are non explicit and it is difficult to understand how they behave with $a$ and $C$. 

More puzzling even is the fact that if the posterior distribution is extremely concentrated, as would happen in the
context of large data sets, the Kullback-Leibler neighbourhood, for a fixed radius $C$, will be very small giving a
(probably) fake impression of robustness. To illustrate this, consider the case where  the posterior distribution is
close to $\mathcal N(\hat \theta, v/n) $ with $n$ large (in other words the posterior distribution verifies a 
Bernstein-von Mises theorem). To understand the impact on $\lambda_a(C ) $ consider also the case where $L_a(\theta)  =
(a-\theta)^2 $. Then $ \pi_{a,C}^{\mbox{sup}} \approx \mathcal N( \mu_n, v_n)$ with $ \mu_n = (\hat \theta  n/v - 2
\lambda_a(C )  a )/(n/v - 2 \lambda_a(C ) )$ and $v_n =  (n/v - 2 \lambda_a(C ) )^{-1}$ 
and it is easy to see that when $\sqrt{n}|a - \hat \theta| \gg 1$, 
$$\lambda_a ( C ) \approx  \sqrt{ \frac{ n C }{ 2 }} |a - \hat \theta |^{-1}$$
while if $a = \hat \theta $ , $ \lambda_a(C  )  \approx \sqrt{C} n /v$. In the latter case  
\begin{equation*}
\psi_{(a)}^{\mbox{sup}} ( C ) \approx \frac{ v }{ n } \left( 1 - u\right)^{-1} , \quad 2C = \frac{ u }{ 1 - u} + \log (1 - u ) 
\end{equation*}
while in the former case, 
\begin{equation*}
\psi_{(a)}^{\mbox{sup}} ( C ) \approx \left( \frac{ v}{ n} + (a-\hat \theta )^2 \right) \left( 1 + \frac{ \sqrt{2C v}}{ \sqrt{n}|a-\hat \theta|} \left( 1 - \frac{ \sqrt{2C v}}{ \sqrt{n}|a-\hat \theta|}\right)^{-1} \right)
\end{equation*}
and the difference between $\psi_{(a)}^{\mbox{sup}} ( C ) $ and $\psi_{(a)}^{\mbox{inf}} ( C ) $ is of the same order as
the risk under $\pi_I$ when $a=\hat \theta$ and is of smaller order than the risk when $a\neq \hat \theta$. Unless
$C$ becomes quite large, the decision $a$ that minimises $\psi_{(a)}^{\mbox{sup}} ( C ) $ is approximately $\hat \theta$.
This behaviour does suggest that the posterior leads to a robust inference, while it could well be that the model is strongly
mis-specified. Such a difficulty appears to be a consequence of using Kullback-Leibler neighbourhoods since $L_1$ neighbourhoods
might have led to a different  behaviour, although this is far from certain.
 

This naturally drives us to question the relevance of the coherence requirement of Theorem 4.2. While the coherence
requirement was quite natural in \citet{bissiri:holmes:walker:2013}, towards building the likelihood, here it is 
unclear why the same result should be obtained directly or after having first observed a subset of observations $x_{(1)}$ and
constructed a least favourable prior $\pi_{a,C}^{\mbox{sup}}(x_{(1)},a)$, prior to observing the rest  of the observations.
We feel that $\pi_{a,C}^{\mbox{sup}}(x,a)$ solely exists after observing $x$ and that it solely makes sense in this very
context. We cannot fathom an equivalent to this result in standard decision-theoretical point estimation. This
is presumably due to the fact that, for a continuum of actions, there is not much appeal in considering a continuum of priors. 
It is however difficult to
evaluate the relevance or importance of this coherence constraint. Once again, generalising to a continuous action space
seems delicate without further guidance. \footnote{  
Note also that the case when the decision $a$ is the parameter $\theta$ itself or rather its estimate
$\hat{\theta}$ as in Section 4.2 remains a puzzle to us as the notation $\theta$ in this section
seems to be used with this double meaning, which makes the outcome questionable.}

The advantage of using the Kullback-Leibler divergence  is that it is  mathematically convenient up to a certain point, but it does not necessarily make sense
from a statistical viewpoint: e.g., is there a model likelihood {\em and} a prior associated with {\em every single}
distribution contained within this neighbourhood? Presumably so, given that this only has to hold for a fixed value of
the sample $x$. And is the least favourable distribution a true posterior for all samples $x$? Presumably not, unless one
accepts ``data-dependent priors".  It may prove to be the most delicate aspect of the paper, namely
that the calibration of the evaluation is highly dependent on how far from the reference value we allow the posteriors
to drift and that we have no clear idea of the meaning of the resulting neighbourhood at the end of Section 4.1. It
seems to us that one of the reasons it is so difficult to calibrate $C$ is that Kullback-Leibler neighbourhoods are rather
abstract objects. Thus one cannot resort to intuition or subjective knowledge in the calibration process.  The discussion by the
authors in the conclusion of this paper broaches upon this very point. In some specific settings, it should be possible to create neighbourhoods by looking at
some quantity of interest and setting bounds or limits on the posterior values or variability for this quantity. An
alternative would be to resort to an ABC \citep{marin:pudlo:robert:ryder:2011} type of robustness where only posteriors
that have the ability to predict the actual data (or the original optimal decision) are allowed within the corresponding
neighbourhood. While this approach is only vaguely defined, and may be delicate to implement in practice, it carries a
most natural kind of proximity.

As a side remark, we appreciate the notion of evaluating the impact of single observations on the inference, as it
creates a decision-based outlier and leverage assessment perspective, but we remain unsure as to what one can conclude
about the appropriateness of the model from the divergence measure, the more because it looks terribly similar to an
harmonic mean estimator \citep{chopin:robert:2007c}!

In connection with this remark, and even before we reached the short section 4.2.3, the introduction of $\Omega$-admissible
actions reminded us of $\Gamma$-minimax procedures mostly studied in the 1980's and early 1990's. This notion did not
attract a large flock of followers at the time, because it is quite delicate to figure out whether or not a procedure is
$\Gamma$-minimax.  However, once we read Section 4.2.3, we got somewhat confused as we could not see how the action
enters the choice of the minimax prior.
Nor why it should.\footnote{We note that the current paper somewhat extends the original $\Gamma$-minimax problem by considering
a collection of posterior rather than priors. While this makes complete sense from a conditioning perspective, it does not
necessarily lead to coherent answers since the least favourable priors are then data dependent.} Clearly,
$\Omega$-admissibility as defined here is a much more convoluted notion that not only involves the range of
acceptable priors but also a new least favourable prior and the optimal decision under the original prior, which somewhat
constitutes the centre of the Kullback-Leibler ball.\footnote{In computational terms, using the centre of the ball for importance
sampling [p.17] may be fraught with danger as the Kullback proximity does not guarantee tail behaviour and hence finite
variance. Hence, using the importance weights to calibrate the Kullback neighbourhood cannot be suggested without
further assessment.} The various proposals of Section 4.2 highlight the links between the proposed approach and other
proposals for robust inference, but they also suggest that the calibration of $C$ can only be achieved on a case by case basis. 

\section{Turning to non-parametric bricks} 

The non-parametric approach via (e.g.) Dirichlet priors is somewhat of an expected if welcomed thread. For one thing, it 
sounds more genuine for inferential purposes, when compared with the call to Kullback-Leibler neighbourhoods. This is
particularly compelling when considering a candidate posterior as the functional parameter of the Dirichlet (hyper)
prior---although the denomination of ``prior" clashes with the fact
that the Dirichlet process is centred at the posterior $\pi_I$---.   Looking at the variability or
distribution of the loss under a Dirichlet process centred at the posterior distribution constitutes quite an
interesting and elegant proposal, but we wonder about the type of \textit{neighbourhoods} of the posterior distributions
thus induced. (We also appreciate the Monte Carlo convenience offered by the Dirichlet process representation described in
Section 4.3.1) 

How do these "neighbourhoods" compare to the Kullback-Leibler neighbourhoods of the first part? 

Indeed Bayesian non-parametrics suffer from the drawback that functional priors are almost irremediably
concentrated on very small regions of the functional space and thus do not necessarily reflect much of a range of possible
posteriors. 
Obviously, in a
nonparametric framework, the notion of neighbourhood is much more loose than in the parametric case. One can envision 
them as soft versus hard neighbourhoods by analogy to soft versus hard thresholding. While the authors state in
Section 4.3.2 that an action optimal under the functional parameter of the Dirichlet prior will remain optimal (in
expectation) under the random measure distributed from this Dirichlet, this may constitute a second-tier property in
that it only stands in expectation.

We find the proposition of studying the probability of changing the optimal decision particularly relevant,  in
particular in the context of discrete and finite decision spaces.  In fact, we are not quite sure of what happens or would happen
in the context of continuous
decisions. Indeed, in this latter case, $L_a(\theta)$ also follows a Dirichlet process centred at the posterior
distribution of $L_a(\theta)$. Hence, if $q_\tau$ is the $\tau$-th quantile of $L_a(\theta)$ under the posterior
$\pi_I$, then $Q( L_a(\theta)\leq q_\tau) \sim {\rm Beta}( \alpha ( 1 - \tau ) , \alpha \tau ) $, which depends only on
$\tau $ and on the mass of the Dirichlet process $\alpha$. There is no dependence whatsoever on $\pi_I$, which seems to
imply it presents little interest. Thus, we wonder which relevant quantities could be produced in continuous
settings towards a better understanding of the variability of the loss function under deviations from the posterior.

\section{Conclusion: The game is afoot!}

While the authors have already uncovered several interesting new avenues for exploring Bayesian robustness, we want to
point out yet another avenue associated with the notion of penalised complexity (PC) priors, as proposed by 
\cite{simpson:etal:2014}, In fact, we think penalised complexity is eminently
relevant to the robustness issue as this perspective tackles the prior specification side covered in Section 2.2. The
starting point for \cite{simpson:etal:2014} modelling is a base model, out of which possibly robust extensions can be
constructed. While there is no automated derivation of the base model, it corresponds to the operational model mentioned
in the current paper. \cite{simpson:etal:2014} further rely on a functional distance from the base and make the quite
natural proposal of setting a prior on the distance from the base. Although there is no decision-theoretic aspect to be
found in this proposal, an extension in this direction is certainly feasible.

In conclusion, we commend the authors for this foray into Bayesian robustess and for producing such a novel perspective.
Formalising this aspect of Bayesian analysis is absolutely essential for methodological and practical purposes, even
when not required by foundational arguments. We also recognise that the proposals made in the paper are mostly exploratory, 
rather than directive, primarily aiming at representing the variability of the
Bayesian output when some of its components are uncertain or misspecified. Once again, this represents an important
step in the rational and objective evaluation of Bayesian procedures and we congratulate the authors for initiating such
a path. Some of the proposals made therein may require further investigation in terms of Bayesian coherence, but 
they open a different perspective on how to envision Bayesian decision making from a braoder viewpoint. 

\section*{Acknowledgements}
The main bulk of those comments were written during the Bayesian week workshop held at CIRM, Luminy, France, March 1-4, 2016, in
a most supportive and serene working environment. We thank the organisers of this meeting for this
great opportunity. We are also grateful to Peter Green for his remarkable patience and to the editorial team for helpful
suggestions towards the final version of this discussion.

\hyphenation{Post-Script Sprin-ger}

\end{document}